\documentclass{article}

\begin{document}
\input epsf

\def\be{\begin{equation}}
\def\ee{\end{equation}}
\def\bea{\begin{eqnarray}}
\def\eea{\end{eqnarray}}
\def\phii{\phi_{i}}
\def\phij{\phi_{j}}
\def\phihati{\hat{\phi_{i}}}
\def\phihatj{\hat{\phi_{j}}}
\def\Phii{\Phi_{i}}
\def\Phij{\Phi_{j}}
\def\Phihati{\hat{\Phi_{i}}}
\def\Phihatj{\hat{\Phi_{j}}}
\def\fof{\frac{\delta S}{\delta\Phi_i}|_{\Phi_{i}=\phi_{i}}}
\def\sof{\frac{\delta^{2} S}{\delta\Phi_i \delta\Phi_j}|_{\Phi=\phi}}
\def\dvdphi{\frac{\partial V}{\partial\phii}}
\def\vecx{\underline{x}}
\def\vece{\underline{e}}
\def\veck{\underline{k}}
\def\kdotx{\underline{k}\cdot\underline{x}}
\def\momvol{\frac{d^{3}\underline{k}}{(2\pi)^{3}}}
\def\nn{\nonumber}
\newcommand{\lrb}{\left(}
\newcommand{\rrb}{\right)}
\newcommand{\ba}{\begin{array}}
\newcommand{\ea}{\end{array}}
\def\adagk{a^{\dag}_{k}}
\def\adag-k{a^{\dag}_{-k}}
\def\a-k{a_{-k}}
\def\ak{a_{k}}
\def\ddagk{d^{\dag}_{k}}
\def\ddag-k{d^{\dag}_{-k}}
\def\d-k{d_{-k}}
\def\dk{d_{k}}

\def\Bdagk{A^{\dag}_{\theta(k)}}
\def\Bdag-k{A^{\dag}_{\theta(k)}}
\def\B-k{A_{\theta(-k)}}
\def\Bk{A_{\theta(k)}}

\def\Ck{D_{\theta(k)}}
\def\Cdagk{D^{\dag}_{\theta(k)}}
\def\C-k{D_{\theta(-k)}}
\def\Cdag-k{D^{\dag}_{\theta(k)}}

\def\Fk{F_{k}}
\def\Fdagk{F^{\dag}_{k}}
\def\F-k{F_{k}}
\def\Fdag-k{F^{\dag}_{-k}}

\def\Adagk{A^{\dag}_{k}}
\def\Adag-k{A^{\dag}_{-k}}
\def\A-k{A_{-k}}
\def\Ak{A_{k}}
\def\Ddagk{D^{\dag}_{k}}
\def\Ddag-k{D^{\dag}_{-k}}
\def\Gdagk{G^{\dag}_{k}}
\def\Gdag-k{G^{\dag}_{-k}}
\def\G-k{G_{-k}}
\def\Gk{G_{k}}
\def\omsumpi{\lrb\frac{\Omega_{\pi}}{\omega_{\pi}}+\frac{\omega_{\pi}}{\Omega_{\pi}}\rrb}
\def\omsumsigma{\lrb\frac{\Omega_{\Sigma}}{\omega_{\Sigma}}+\frac{\omega_{\Sigma}}{\Omega_{\Sigma}}\rrb}
\def\omdifpi{\lrb\frac{\Omega_{\pi}}{\omega_{\pi}}-\frac{\omega_{\pi}}{\Omega_{\pi}}\rrb}
\def\omdifsigma{\lrb\frac{\Omega_{\Sigma}}{\omega_{\Sigma}}-\frac{\omega_{\Sigma}}{\Omega_{\Sigma}}\rrb}
\def\bmat{\lrb\begin{array}{c}}
\def\emat{\end{array}\rrb}
\def\phii{\phi_{i}}
\def\lb{\label}

\title{Charged vs. Neutral particle creation in expanding Universes: A Quantum Field Theoretic
treatment.}

\author{{\bf B.Bambah}\footnote{Presented by B.Bambah at the {\it International Congress
of Mathematical Physics (ICMP2003)}, Lisbon Portugal, July 2003.}
 \\ School of Physics , University of
  Hyderabad, \\ Hyderabad-500 046,India.\\
  {\bf C.Mukku}\\ Department of
  Mathematics\\ and\\ Department of Computer Science \& Applications, \\Panjab University, Chandigarh-160014,India }
\maketitle

\begin{abstract}
  A complete quantum field theoretic study of charged and neutral particle
  creation in a rapidly/adiabatically expanding Friedman-Robertson-Walker metric
  for an O(4) scalar field theory with quartic interactions (admitting a phase transition) is given.
  Quantization is carried out by inclusion of quantum fluctuations.
  We show that the quantized Hamiltonian admits an su(1,1) invariance.
  The squeezing transformation diagonalizes the Hamiltonian and
  shows that the dynamical states are squeezed states. Allowing
  for different forms of the expansion parameter, we show how the
  neutral and charged particle production rates change as the
  expansion is rapid or adiabatic.
  The effects of the expansion rate versus the symmetry
  restoration rate on the squeezing parameter is shown.
\end{abstract}

\section{Introduction}
 In the cosmological context, early studies on the creation of
particles in curved spacetimes have concentrated on neutral scalar
fields \cite{birrell}, \cite{parker}. While these studies led to
an understanding of relationships between particle physics in the
early universe and cosmology, the inclusion of spontaneous
symmetry breaking coupled with the HEP scenario for the early
universe has led to the recent resurgence in inflation being
driven by the fluctuations in quantum fields undergoing symmetry
breaking (preheating) \cite{linde,brand}. The basic methodologies
used for such studies has largely been the Landau -Ginzburg model
\cite{boya}. In this paper we advocate the use of a Hamiltonian
generated through an O(4) linear sigma model with quartic
interactions for such studies. The Hamiltonian is a generic
Hamiltonian applicable to many domains \cite{qftdcc}. In this
paper we show how it can be used to determine the production rates
of charged vs. neutral particles in an expanding metric. We also
show how squeezed states are the correct dynamical states for the
fields in expanding spacetimes. We hope to show in a later
communication, how these states may be used to study the
"tachyonic" inflationary cosmological models of current interest.
\section{The Hamiltonian.}
The growth of long wavelength fluctuations during breaking of
symmetries has been studied recently for the case of a single
scalar field with quartic interactions \cite{linde}. While they
have used Lattice simulations, we advocate a model Hamiltonian for
quantum fluctuations of charged and neutral scalar fields which we
shall derive here. Our starting point is the O(4) linear sigma
model with an action given by: \be S=\int d^{3}x dt a(t)^{3}
(\frac{1}{2} \dot{\Phi}_{i}^{2}
-\frac{1}{2a^{2}}(\nabla\Phi_i)^{2} -\frac{1}{2} m^{2}\Phi_{i}^{2}
-\frac{\lambda}{4\!}(\Phi_{i}^{2}-v^2)^2 ), \ee with
 \be \Phi_i=
 \lrb \ba{c}\Phi_1 \\  \Phi_2
\\  \Phi_3 \\ \Phi_4 \ea \rrb
\ee
 where $\Phi_i$, $i=1..4$ are real scalar fields.
The background spacetime is taken to be a FRW model: \be
ds^{2}=dt^{2}-a(t)^{2} d\vec{x}^{2}, \ee where $a(t)$ is the
expansion parameter. We now use a background field analysis to
study the quantum effects. Assume $\Phii$ has a background
classical component $\phii$ which satisfies the classical
equations of motion:
\be
\frac{\delta S}{\delta\Phii}|_{\Phii=\phii} = 0. \ee Treat the
quantum field $\phihati$ as  fluctuation around classical solution:
\be
\Phi_{i} \longrightarrow \phii + \phihati \ee
since  $\phii$ satisfies the classical equations of motion,
\be
S=S[\phii ] + \frac{1}{2}\phihati\sof\phihatj +\cdots. \ee We
shall deal with a quadratic fluctuation action given simply by:
\be S_{2}=\frac{1}{2}\phihati\sof\phihatj . \ee For the particular
scalar field action given above, assuming all fields vanish at
infinity (asymptotically flat metrics), \be \frac{\delta
S}{\delta\Phij} = -\partial_{\mu}(a^3
g^{\mu\nu}\partial_{\nu}\Phij)-a^3 m^2\Phij -a^3 \frac{\delta
V}{\delta\Phij}. \ee Imposing the classical equations of motion,
we find that \be
\partial_{\mu}(a^3 g^{\mu\nu}\partial_{\nu}\phii)+a^3 m^2\phii +a^3 \frac{\partial V}{\partial\phii} = 0
\ee where $\frac{\partial V}{\partial\phii}\equiv\frac{\partial
V}{\partial\Phii}|_{\phii}$. The equations of motion in this
metric are \be 3\frac{\dot{a}}{a}\dot{\phii} +\dot{\phii}^{2}
-\frac{1}{a^{2}}\underline{\nabla}^{2}\phii +m^{2}\phii +\dvdphi =
0. \ee Since we are interested in the dynamics of the fluctuation
field, we shall treat the fluctuation field in $S_{2}$ as a
classical field and $S_{2}$ itself as the classical action for its
dynamics.  We  define a Lagrangian density for studying the
dynamics of the fluctuations, $\cal{L}$, as follows:
\be
{\cal{L}}=\frac{a^3}{2}(\phihati^{2}-\frac{1}{a^{2}}(\underline{\nabla}\phihati)^{2}
-m^{2}\phihati^{2} -\phihati\frac{\partial^{2}
V}{\partial\Phii\partial\Phij}|_{\phi}\phihatj). \ee Carrying out
a Legendre transformation, it is easy to write down the
Hamiltonian density \be {\cal{H}}={\frac{1}{2a^{3}}\hat{p_{i}}^{2}
+\frac{a}{2}(\underline{\nabla}\phihati)^{2}+\frac{a^{3}m^{2}}{2}\phihati^{2}
 +\frac{a^3}{2}(\phihati\frac{\partial^{2}
V}{\partial\Phii\partial\Phij}|_{\phi}\phihatj)} -\frac{\lambda
a^3 v^2}{2}\phihati^2\ee where \be \hat{p_{i}}=\frac{\delta
\it{L}}{\delta \dot{\phihati}}=a^{3}\dot{\phihati}. \ee
 We also have
\begin{equation}
\frac{\partial^2 V}{\partial\Phi_i \partial\Phi_j}|_\phi =2
\lambda \phi_i\phi_j+\lambda(\phi_k^2-v^2) \delta_{ij}.
\end{equation}
Assume that the fluctuation field $\Phi_i$ decomposes into its
constituents as:\\
\begin{equation}
\hat{\phi}=<\Phi>-\phi=\left(\begin{array}{c}\phi_1\\ \phi_2\\
\phi_0\\ \Sigma\end{array}\right).
\end{equation}
The physical fields are defined so that
\begin{equation}
  \phi_+=\frac{1}{\sqrt{2}}(\phi_1+i\phi_2);\hspace{2cm}
  \phi_-=\frac{1}{\sqrt{2}}(\phi_1-i\phi_2).
\end{equation}
Analogously, we define the classical background fields as:
\begin{equation}
  v_+=\frac{1}{\sqrt{2}}(v_1+iv_2);\hspace{2cm}
  v_-=\frac{1}{\sqrt{2}}(v_1-iv_2),\\
\end{equation}
following the identification:
\begin{equation}
\phi=\left(\begin{array}{c}v_1\\ v_2\\ v_3=v\\
\sigma\end{array}\right)\equiv <\Phi>.
\end{equation}
It is easy to see that a Legendre transformation provides a
Hamiltonian given as: \be H=H_{neutral}+H_{charged}+H_{mixed}, \ee
where  \bea H_{neutral}&=&\int d^3xdta^3 \{
\frac{(p_{0\phi_0}^2)}{2a^6}+\frac{1}{2a^2}(\nabla\phi_0^2)+\frac{1}{2}(m_{\phi}^2)(\phi_0)^2+\frac{(\Omega_{\phi}^2-\omega_{\phi}^2)}{2a^6}\phi_0^2
\nn \\
&+&\frac{(p_{0\Sigma}^2)}{2a^6}+\frac{1}{2a^2}(\nabla\Sigma^2)+\frac{1}{2}(m_{\Sigma}^2)(\Sigma)^2+\frac{1}{2a^6}(\Omega_{\Sigma}^2-\omega_{\Sigma}^2)(\Sigma)^2\}
\eea \bea H_{charged}&=&\int d^3x dt a^3 (\frac{1}{a^6}
\{(p_{0+}p_{0-})\nn \\
&+&\frac{1}{a^2}(\nabla\phi_-)(\nabla\phi_+)+(m_\phi^2)(\phi_+\phi_-)+\frac{1}{a^6}(\Omega_{\phi_{\pm}}^2-\omega_{\phi_{\pm}}^2)(\phi_+\phi_-)\}
\eea \bea
 H_{mixed}&=&\int d^3x
a^3dt\{\lambda(v_{+}^2\phi_{-}^2+v_{-}^{2}\phi_{+}^2)
 \nn \\
&+&(2\lambda)(v_+v_3\phi_-\phi_0+ v_-v_3\phi_+\phi_0+\sigma
v_3\Sigma\phi_0+v_+\sigma\phi_-\Sigma+\sigma v_-\Sigma\phi_+)\}.
\eea Where we have also put \be
\frac{\Omega^2_{\phi_0}-\omega^2_{\phi_0}}{a^6}=\lambda[2v_3^2];
\frac{\Omega^2_{\phi
{\pm}}-\omega^2_{\phi_{\pm}}}{a^6}=\lambda[2v_+v_-];
\frac{\Omega^2_{\Sigma}-\omega^2_{\Sigma}}{a^6}=\lambda[2\sigma^2]
\ee and \be 2v_+v_-+v_3^2+\sigma^2=v^2. \ee This is the most
general Hamiltonian for the O(4) scalar field system, now broken
up into a charged scalar field and two neutral scalar fields in
the background field formalism. The background field can now be
parametrized through three angles:\be \phii=\lrb \ba{c} v
Cos(\rho)Sin(\theta)Sin(\alpha)\\vCos(\rho)Sin(\theta)Cos(\alpha)\\vSin(\rho)Sin(\theta)\\vCos(\theta)\ea
\rrb.\ee Here we shall simplify the parametrization of the
background field to two angles, $\theta$ and $\rho$ by letting
$\alpha=\frac{\pi}{4}$: then,
$v\pm=\frac{v}{\sqrt{2}}Cos(\rho)Sin(\theta)$,\hspace{0.5cm}$v_{3}=v
Sin(\rho)Sin(\theta)$,\hspace{0.5cm} and
\hspace{0.5cm}$\sigma=vCos(\theta)$. Canonical quantization gives
the mode decomposed Hamiltonian:

 \bea
 H_{neutral}=\int \momvol
\frac{1}{2}\{
\frac{\omega_{\phi}}{a^3}(a_k^{\dag}a_k+a_ka^{\dag}_k)  \nn \\
+\frac{\omega_{\phi}}{2a^3}(\frac{\Omega_{\phi}^2}{\omega_{\phi}^2}-1)(a_k^{\dag}a_k+a_ka_k^{\dag}+a_{-k}a_{k}+a_{-k}^{\dag}a_{k}^{\dag})
 \nn \\
 +\frac{\omega_{\Sigma}}{a^3}(
 d_k^{\dag}d_k+d_kd^{\dag}_k)+\frac{\omega_{\Sigma}}{2a^3}(\frac{\Omega_\Sigma^2}{\omega_{\Sigma}^2}-1)(d_k^{\dag}d_k+d_kd_k^{\dag}+d_{-k}d_{k}+d_{-k}^{\dag}d_{k}^{\dag})\}
\eea
\bea
 H_{charged}=\int \momvol
\{ \frac{\omega_{\phi}}{a^3}(b_k^{\dag}b_k+c_kc^{\dag}_k)  \nn \\
+\frac{\omega_{\phi}}{2a^3}(\frac{\Omega_{\phi_\pm}^2}{\omega_{\phi}^2}-1)(b_k^{\dag}b_k+c_kc_k^{\dag}+b_{-k}c_{k}+c_{-k}^{\dag}b_{k}^{\dag})\}
\eea \bea
 H_{mixed}=\int \momvol
\{ \frac{\lambda a^3v^2cos^2(\rho)sin^2(\theta)}{4\omega_\phi}(
 b_k b_{-k}+b_{k}c^{\dag}_k
 \nn \\+c^{\dag}_k b_{k}+c_k c_{-k}+c^{\dag}_k
 c^{\dag}_{-k}+c_kb^{\dag}_k+b^{\dag}_{k}c_{k}+b^{\dag}_{k}b^{\dag}_{-k})
 \nn \\
+ \frac{\lambda
a^3v^2cos(\rho)sin(\rho)sin^2(\theta)}{\sqrt{\omega_{\Sigma}\omega_\phi}}
\nn\\
\lrb b_k a_{-k}+b_{k}a^{\dag}_k+c^{\dag}_k a_{k}+c_k
a_{-k}+c^{\dag}_k
a^{\dag}_{-k}+c_ka^{\dag}_k+b^{\dag}_{k}a_{k}+b^{\dag}_{k}a^{\dag}_{-k}\rrb \nn \\
+ \frac{\lambda
a^3v^2sin(\rho)sin(\theta)cos(\theta)}{\sqrt{\omega_\phi\omega_{\Sigma}}}\lrb
 d_k a_{-k}+d_{k}a^{\dag}_k+d^{\dag}_k a_{k}+d^{\dag}_k
 a^{\dag}_{-k}\rrb
 \nn \\
+ \frac{\lambda
a^3v^2cos(\rho)sin(\theta)cos(\theta)}{\sqrt{\omega_{\phi}\omega_{\Sigma}}}
\nn \\ \lrb
 b_k d_{-k}+b_{k}d^{\dag}_k+c^{\dag}_k d_{k}+c_k d_{-k}+c^{\dag}_k d^{\dag}_{-k}+c_kd^{\dag}_k+b^{\dag}_{k}d_{k}+b^{\dag}_{k}d^{\dag}_{-k}\rrb
 \}\eea
 where \be
\frac{\omega^{2}_{\phi}(k)}{a^{6}}\equiv\frac{\omega^{2}_{\phi_0}(k)}{a^{6}}=\frac{\omega^{2}_{\phi_\pm}(k)}{a^{6}}=(m_{\phi}^{2}+\frac{\veck^{2}}{a^{2}});\;\;
\frac{\omega^{2}_{\Sigma}(k)}{a^{6}}=(m_{\Sigma}^{2}+\frac{\veck^{2}}{a^{2}})
\ee and \be
\frac{\Omega_{\phi}^{2}(k)}{a^6}=\frac{k^2}{a^2}+m_{\phi}^{2}+\lambda
(v^{2}+2v_{3}^{2});\;\;
\frac{\Omega_{\phi_\pm}^{2}(k)}{a^6}=\frac{k^2}{a^2}+m_{\phi_\pm}^{2}+\lambda(v^{2}+2v_{+}v_{-}).\ee
An important point to note here is that while $\Omega_{\phi}(k)$
and $\omega_{\phi}(k)$ are momentum dependent, for ease of
notation we will drop the k dependence for further calculations
and revive it when necessary. Clearly, there are two interesting
cases to be considered here: $\theta=0$ and $\rho=\frac{\pi}{2}$.
For the purposes of this study, we shall only consider the first
case, $\theta=0$. Its easy to see that $H$ reduces to
 \bea
 H=\int \momvol
\{ \frac{\omega_{\phi}}{2a^3}(a_k^{\dag}a_k+a_ka^{\dag}_k) \nn \\
+\frac{\omega_{\phi}}{4a^3}(\frac{\Omega_{\phi}^2}{\omega_{\phi}^2}-1)(a_k^{\dag}a_k+a_ka_k^{\dag}+a_{-k}a_{k}+a_{-k}^{\dag}a_{k}^{\dag})
+\frac{\omega_{\Sigma}}{2a^3}(
 d_k^{\dag}d_k+d_kd^{\dag}_k)
\nn \\+\frac{\omega_{\Sigma}}{4
 a^3}(\frac{\Omega_\Sigma^2}{\omega_{\Sigma}^2}-1)(d_k^{\dag}d_k+d_kd_k^{\dag}+d_{-k}d_{k}+d_{-k}^{\dag}d_{k}^{\dag})\}\nn
 \\+\{\frac{\omega_{\phi}}{a^3}(b_k^{\dag}b_k+c_kc^{\dag}_k)+\frac{\omega_{\phi}}{2a^3}(\frac{\Omega_{\phi}^2}{\omega_{\phi}^2}-1)(b_k^{\dag}b_k+c_kc_k^{\dag}+b_{-k}c_{k}+c_{-k}^{\dag}b_{k}^{\dag})\}
 \eea

This Hamiltonian has an $su(1,1)$ symmetry and can be diagonalized
by a series of Bogolyubov transformations: In the neutral sector,

\be A_{k}(t,r)=\mu(r,t) a_{k}+\nu(r,t)
a^{\dag}_{-k}=U^{-1}(r,t)a_{k}U(r,t).\ee Similarly for the sigma
field, $D_{k}(t,r')$ and $D^{\dag}_{k}(t,r')$ are defined in
analogy with the definition of $A_{k}(t,r)$ and
$A^{\dag}_{k}(t,r)$ with the d's replacing the a's. For the
charged sector \be C_{k}(t,r)=\mu c_{k} +\nu
b^{\dag}_{-k}=U^{-1}(r,t)c_{k}U(r,t)  \ee
 \be
B_{k}(t,r)=\mu c_{-k} +\nu b^{\dag}_{k}=U^{-1}(r,t)b_{k}U(r,t) \ee
where\be
\mu=Cosh(r)=\sqrt{\frac{1}{2}[(\frac{\Omega_\phi}{\omega_\phi}+\frac{\omega_\phi}{\Omega_\phi})+1]}
\; \;\;\;\;\
\nu=Sinh(r)=\sqrt{\frac{1}{2}[(\frac{\Omega_\phi}{\omega_\phi}+\frac{\omega_\phi}{\Omega_\phi})-1]}
\ee and $r$ is the squeezing parameter, satisfying
$\mu^{2}-\nu^{2} = 1$ as required for a squeezing transformation.
The complete unitary matrix for the squeezing transformation may
be written down as \be U(r,t)=e^{\int \momvol
r(k,t)\{(a^{\dag}_{k}a^{\dag}_{-k}-a_{k}a_{-k})+(d^{\dag}_{k}d^{\dag}_{-k}-d_{k}d_{-k})+(c_{k}b_{-k}+b_{k}c_{-k})-(c^{\dag}_{k}b^{\dag}_{-k}+b^{\dag}_{k}c^{\dag}_{-k})\}}
\ee Collecting all our results for the neutral and charged
sectors, the total diagonalized Hamiltonian is written in terms of
various creation and annihilation operators as: \be H=\int
\momvol\frac{1}{2a^3}\{ \Omega_{\phi}\{(A^{\dag}_{k} A_{k} +
\frac{1}{2}) +( C^{\dag}_{k} C_{k} + B^{\dag}_{k}
B_{k}+1)\}+\Omega_{\Sigma}(D^{\dag}_{k}D_{k}+\frac{1}{2})\}. \ee
 Since the $\Sigma$ field decouples, we drop all terms
associated with it whenever it is not essential to our arguments.
The total dynamical Hamiltonian for the neutral and charged scalar
fields in terms of the creation and annihilation operators
($a,a^{\dag},c,c^{\dag},b$ and $b^{\dag}$)simplifies to: \bea
H_{0}&=&\int
\momvol\frac{\Omega_{\phi}}{a^3}\{2(\mu^{2}+\nu^{2})(c^{\dag}_{k}c_{k}
+b^{\dag}_{k}b_{k}+1) \nn \\ &
+&\mu\nu\{(c_{k}b_{-k}+b_{k}c_{-k})+(c^{\dag}_{k}b^{\dag}_{-k}+b^{\dag}_{k}c^{\dag}_{-k})\}\nn
\\
&+&\frac{\Omega_{\phi}}{a^3}\{(\mu^{2}+\nu^{2})(a^{\dag}_{k}a_{k}+1)+\nu\mu(a^{\dag}_{-k}a^{\dag}_{k}+a_{k}a_{-k})\}.
\eea We have completed the derivation of a Hamiltonian describing
the dynamics of neutral and charged scalar fields with a symmetry
breaking parameter in an expanding FRW metric. We shall see that
these two parameters allow us to examine the amplification of
charged and neutral modes as they are varied, both in a sudden
quench and adiabatically. We have already seen in [\cite{qftdcc}]
that this Hamiltonian can be used to describe the formation and
decay of the disoriented chiral condensate (DCC). The generic
structure of the Hamiltonian allows us to apply its dynamics to
the inflationary studies in cosmology. We also believe it provides
an excellent framework for analyses of the Bose-Einstein
condensate (BEC)- this is left for a later communication.

\section{Evolution of the fluctuations}
We note first that the sigma field decouples in this particular
Hamiltonian and therefore it can be analyzed independently of the
$\phi_0,\phi_\pm$ fields. We write the total dynamical Hamiltonian
in terms of the creation and annihilation operators
($a,a^{\dag},c,c^{\dag},b $ and $b^{\dag}$)and the squeezing
parameters as: \bea H&=&\int
\momvol\frac{\Omega_{\phi}}{a^3}\{2(\mu^{2}+\nu^{2})\{c^{\dag}_{k}c_{k}
+b^{\dag}_{k}b_{k}+1\}\nn \\ &+&\mu\nu\{(c_{k}b_{-k}+b_{k}c_{-k})
+(c^{\dag}_{k}b^{\dag}_{-k}+b^{\dag}_{k}c^{\dag}_{-k})\} \nn
\\+&&\frac{\Omega_{\phi}}{a^3}\{(\mu^{2}+\nu^{2})\{a^{\dag}_{k}a_{k}+1\}+(\nu\mu)\{a^{\dag}_{-k}a^{\dag}_{k}+a_{k}a_{-k}\}\}.
\eea Defining the following bilinear operators: \bea
 {{{\cal{D}}}}&=&a_{k}a_{-k}+b_{k}c_{-k}+c_{k}b_{-k}=K_1^- +K_2^-
 +K_3^-
 \nn\\
 {{{\cal{D}}}}^{\dagger}&=&a^{\dag}_{-k}a^{\dag}_{k}+c^{\dag}_{-k}b^{\dag}_{k}+b^{\dag}_{-k}c^{\dag}_{k}=K_1^+ +K_2^+
 +K_3^+
 \nn\\
 N&=&\frac{1}{2}\{a^{\dag}_{k}a_{k}+a^{\dag}_{-k}a_{-k}+b^{\dag}_{k}b_{k}+b^{\dag}_{-k}b_{-k}+c^{\dag}_{k}c_{k}+c^{\dag}_{-k}c_{-k}+3\}\nn \\ &=&K_1^0+K_2^0+K_3^0
 \nn\eea it is easy to see that they satisfy an $su(1,1)$ algebra \be
[N,{{{\cal{D}}}}]=-{{{\cal{D}}}};\;\;\;\;
[N,{{{\cal{D}}}}^{\dag}]={{{\cal{D}}}}^{\dag};\;\;\;
[{{{\cal{D}}}}^{\dag},{{{\cal{D}}}}]=-2N. \nn \ee The $su(1,1)$
invariant Hamiltonian for the $\phi_0,\phi_\pm$fields assumes the
form
\begin{equation}
H=\int \momvol\frac{1}{a^3}
2\Omega_{\phi}(k,t)(\mu^2+\nu^{2})N+2\Omega_{\phi}(k,t)\mu\nu({{\cal{D}}}+{{\cal{D}}}^{\dag})
\end{equation}
The  time dependent evolution equation  is given by
\be
H(t)|\psi(t)>=i\frac{d}{dt}|\psi(t)> \ee The particular $su(1,1)$
structure elucidated above provides us the solution:
 \be |\psi(t)>=e^{\int \momvol
r_k({{\cal{D}}}^{\dag}_{k}-{{\cal{D}}}_{k})}|\psi(0> \lb{wf} \ee
for the evolution of the wave function immediately. Here $r_k$ is
the squeezing parameter related to the physical variables
$\Omega_{\phi}(k,t)$ and $\omega_{\phi}(k)$ through \be
Tanh(2r_k)= \frac{(\frac{\Omega_{\phi}(k,t)}{\omega_\phi})^2-1}
 {(\frac{\Omega_{\phi}(k,t)}{\omega_\phi})^2+1} \ee Where $\Omega_{\phi}(k,t\longrightarrow\infty)=\omega_{\phi}(k)$.
It is assumed that past and future infinity is flat spacetime.
Thus in the evolution of the fluctuations, it is the frequency
changes which bring about squeezing.

The diagonalized Hamiltonian $H_{0}$ can be converted into a
Hamiltonian in terms of quantum fields corresponding to the
operators $A,B,C $ and their adjoints to obtain a purely quadratic
Hamiltonian.

We can, for example, write \be
\frac{\Omega_\phi}{a^3}(\Adagk\Ak+\frac{1}{2})=\frac{\Omega_\phi}{a^3}
(\Adagk\Ak
+\Ak\Adagk)=(\frac{\Omega_\phi}{a^3})^2\Pi_A^2(k,t)+P_{\Pi_{A}}^2(k,t).\ee
Similarly for B and C [\cite{qftdcc}]. The Hamiltonian $H_0$ can
then be written as: \be
H_0(t)=\int\frac{d^3k}{(2\pi)^3}\sum_{i=A,B,C}\frac{1}{2}((\frac{\Omega_{\phi}}{a^3})^2\Pi_{i}^2(k,t)+P_{\Pi_{i}}^2(k,t))\ee

The Schroedinger equation for each momentum mode is simply: \be
H_0(k,t)\psi(k,t)=i\frac{d}{dt}\psi(k,t).\ee If we use the
$\Pi$-representation (co-ordinate space representation) for
$\psi(k,t)$, then, the $su(1,1)$ symmetry of the Hamiltonian tells
us that the solution for $\psi(k,t)$ is just a Gaussian. The
equation satisfied by the wave functions for each mode are then
given by: \be
\ddot\psi_{A}(k,t)+\frac{3\dot{a}}{a}\dot{\psi_{A}}+(\frac{\Omega_{\phi}}{a^3})^{2}(k,t)\psi_{A}(k,t)=0.\ee
(similar equations hold for fields B and C) where \be
(\frac{\Omega_\phi}{a^3})^2(k,t)=(\frac{\veck^2}{a^2})+m_{\phi}^2+\lambda
v^2.\ee The expectation values of the number operator for the
neutral scalar fields for each momentum k is given by: \be
<\psi_{k}(t)|a^{\dag}_{k}a_{k}|\psi_{k}(t)>=Sinh^{2}(r)=<\psi_{k}|A^{\dag}_{k}(t)A_{k}(t)|\psi_{k}>.\ee
An identical expression holds for the charged scalar fields. We
let \be d\eta=a(t)^{-1}dt \ee so that the FRW metric is
transformed into a conformally flat metric:
$ds^2=a(\eta)^2(d\eta^2-d\vecx^2)$. The equations of motion given
above are then transformed into ones that resemble a harmonic
oscillator with time dependent frequencies. We shall write only
the generic form of the above equations: In terms of the scaled
time, $\eta$, we have: \be \psi''+\frac{2 a'}{a}\psi' +(\vec
k^2+(m_{\phi}^2+{\lambda} (<\Phi^2>-v^2)
  )a^2\psi =0\ee
where a prime denotes differential wrt $\eta$.

Lastly, scaling $\psi$: $ \xi=a\psi $ so that the equation
becomes: \be -\xi''+V(\eta)\xi =(k^2+m_{\phi}^2)\xi \ee where \be
V(\eta)=a^{-1}\frac{d^{2}a}{d\eta^2}+m_{\phi}^{2}(1-a^{2})-\lambda
(<\Phi^2>-v^2).\lb{dual}\ee Thus in the symmetry broken stage
$<\Phi^2>=v^2$ \be
V_b(\eta)=a^{-1}\frac{d^{2}a}{d\eta^2}+m_{\phi}^{2}(1-a^{2})
\lb{vb}\ee and in the symmetry restored stage,$<\Phi^2>=0$ \be
V_r(\eta)=a^{-1}\frac{d^{2}a}{d\eta^2}+m_{\phi}^{2}(1-a^{2})+\lambda
a^{2}v^2. \lb{vr}\ee

 These equations can be interpreted in two ways allowing for
simple solutions. Treating $\eta$ as a spatial variable allows us
to treat them as Schroedinger like equations with
$E=\omega_{\phi}^2$. This then allows calculation of the
reflection and transmission coefficients over the ``potential
barrier'' provided by the $V(\eta)$ term. On the other hand, they
are equations for time dependent harmonic oscillators with time
dependent frequencies given by $\Omega_{\phi}^2$ and
$\omega_{\phi}^2$. These two pictures allow the calculation of the
squeezing parameter dependent number operator $N(k)=Sinh^2(r(k))$
involved in the evolution of the neutral and charged quantum
fluctuations in the FRW background.

 We shall now look at the scenario in the FRW metric with
expansion included to show that the enhancement of the low energy
modes and the squeezing parameter are dependent on the rate of the
expansion mechanism by which the symmetry is restored. We have
also seen that to produce substantial squeezing, we require a
quenching. To show this really happens in this case, we have to
compare the situation of a sudden quench with a slow adiabatic
relaxation of the system from the symmetry restored stage  to the
symmetry broken stage.

To examine the amplification of the neutral and charged modes for
the case of a smoothly expanding metric, we shall assume the
expansion parameter $a(\eta)$ to be\cite{birrell}: \be
a(\eta)^2=A+B Tanh[b\eta].\lb{ar}\ee

Here $\frac{1}{b}$ measures the rate of the expansion and we
choose b to be in units of $m_{\phi}$. When viewed as a time
dependent oscillator  equation in an expanding metric we may write
the equation for $\xi$ in the symmetry restored phase as an
oscillator equation \be \xi''+\omega_{\phi}^2\xi-V_r(\eta)\xi =0
\ee and in the symmetry broken stage as \be
\xi''+\omega_{\phi}^2\xi-V_b(\eta)\xi =0.\ee Where
 $V_b(\eta)$and $V_r(\eta)$ are given by (\ref{vr},\ref{vb}) with the particular choice
of the expansion parameter $a(\eta)$ given by (\ref{ar}). Thus the
change in frequency from the restored to the broken stage is given
by \be V_r(\eta)-V_b(\eta)=+\lambda v^2 a(\eta)^2 \ee

Here, we shall choose A and B in eqn.(\ref{ar}) to be 1. The
transmission coefficients across such a barrier are easily looked
up from \cite{birrell}. While the number of particles of mode $k$
is given by: \be N(k)=\frac{(Sinh^2(\frac{\pi\lambda
v^2}{b}))}{Sinh(\frac{\pi (k^2 +m_{\phi}^2)}{b})Sinh(\frac{\pi
(k^2 + m_{\phi}^2 +2 \lambda v^2)}{b})} \ee
 Here b measures the duration of the quench and its effect is exhibited in figure 1.
\begin{figure}[hbp]
\begin{center}
\epsfxsize=7cm\epsfbox{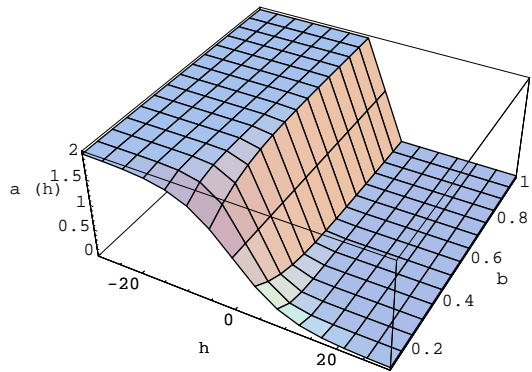}
\end{center}
\caption{Shows
the transition from quench to adiabatic for various values of b. }
\end{figure}

$N(k) vs. k $ is plotted in fig. 2. We see that in the adiabatic
limit of large $b$, $N(k)$ is exponentially suppressed so that
there is no enhancement of low momentum modes.

\begin{figure}[tbp]
\begin{center}
\epsfxsize=7cm\epsfbox{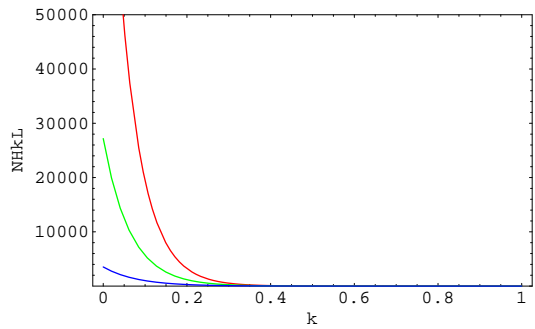} \end{center} \caption{Shows the
variation N(k) with k for values of  b=0.35 (red); 0.4 (green);
0.5 (blue).}
\end{figure}

From the above we conclude that if expansion is rapid we get the
quenched limit, while, if expansion is slow, we get the adiabatic
limit. Since in the squeezed state description $N(k)=Sinh^2r_k$
,where $r_k=r(k)$ is the squeezing parameter , we conclude that in
the quenched limit the squeezing parameter is much greater than in
the adiabatic limit. This demonstrates clearly the connection
between the rate of expansion and squeezing and the enhancement of
charged and neutral particle creation.

Since the amplification is large for the low momentum modes, let
us look at the charged and neutral number particle distributions
for zero momentum. We shall add a subscript $0$ to denote zero
momentum modes so that the distributions are given by:\bea
P_{n_0,n_+,n_-}&=&|<n_{0},n_{+},n_{-}|\psi>|^{2}\nn \\&=&
<n_{0}|e^{r_0(a_{0}^{\dag})^{2}-r_0*a_{0}^{2}}|0>
<n_{+},n_{-}|e^{2r_0 (b_{0}
^{\dag}c_{0}^{\dag}-b_{0}c_{0})}|0>|^{2} \eea defining $S(r_0)$ as
the one mode squeezing operator \be S(r_0)= <n_{0}|e^{r_0
((a_{0}^{\dag})^{2}-a_{0}^{2})}|0> = S_{n_{0},0}. \ee
$S^{tm}(r_0)$ is then the two mode squeezing operator \be
<n_{+},n_{-}|e^{(r_0(b_{0}^{\dag}c_{0}^{\dag}-b_{0}c_{0}))}|0>
=S^{tm}_{n_{+},n_{-},0}. \ee The neutral and charged particle
number distribution is:
\begin{equation}
P_{n_0,n_c}=<S_{n_{0},0}>^{2}<S^{\dag}{}^{m}_{n_{+},n_{-},0,0}>^{2}
\end{equation}
 Writing
$n_{+}=n_{-}=n_{c}$, we get the distribution of charged particles
to be
\begin{equation}
P_{n_{c}}{}=\sum_{n_0}P_{n_0,n_c}=
\frac{(tanh(r_0))^{2n_{c}}}{(cosh(r_0))^{2}}
\end{equation}
and the distribution of neutral particles to be
\begin{equation}
P_{n_{0}}=\sum_{n_c}P_{n_0,n_c}=\frac{n_{0}!(tanh(r_0))^{n_{0}}}{((\frac{n_{0}}{2})!)^{2}cosh(r_0)
2^{n_{0}}}
\end{equation}

 Thus the neutral
and charged particle number distributions are significantly
different as the two types of squeezed states appearing in the
expressions above have different properties. We now illustrate the
effect of squeezing in these two distributions. Figs. 3, and 4
show the difference in the charged and neutral particle
distributions as we vary  the squeezing parameter  from a low
value to  a high value.

\begin{figure}[tbp]
\begin{center}
\epsfxsize=7cm\epsfbox{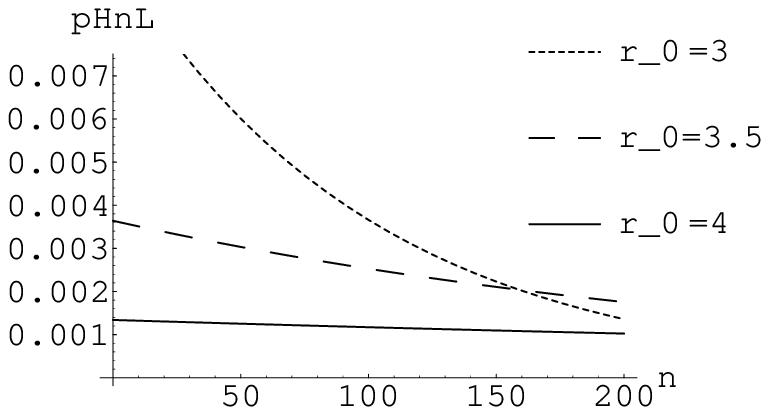} \end{center} \caption{Shows the
variation of $P_{nc}$ with $n$ for the $r_{0}=3,3.5$ and $ 4$}
\end{figure}

\begin{figure}[tbp]
\begin{center}
\epsfxsize=6cm\epsfbox{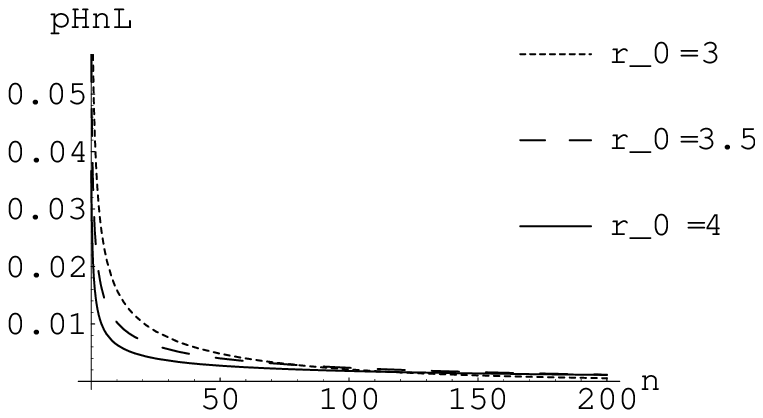} \end{center} \caption{Shows the
variation of $P_{nc}$ with $n$ for the $r_{0}=3,3.5$ and $ 4$}
\end{figure}
  We now illustrate the effect of quenching versus
adiabaticity on these two distributions. Figs. 5 and 6  show the
difference in the charged and neutral particle number
distributions as we vary from the adiabatic limit where the
difference is negligible to the quenched limit where the
difference is significant.
\begin{figure}[htbp]
\begin{center}
\epsfxsize=6cm\epsfbox{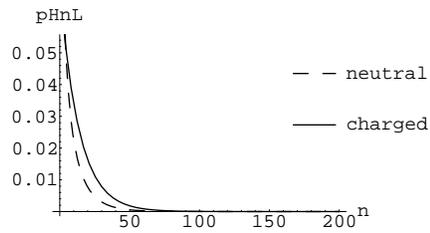}\end{center} \caption{Shows the
variation of $P_{n0}$ (solid line)and $P_{nc}$(dashed line) with
$n$ for the {\it adiabatic} limit ($r_0=2$)}
\end{figure}
\begin{figure}[htbp]
\begin{center}
\epsfxsize=6cm\epsfbox{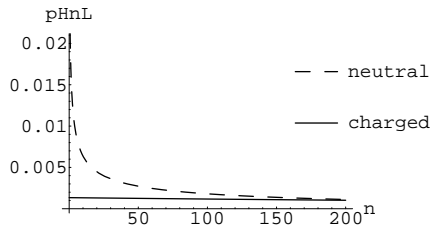}\end{center} \caption{Shows the
variation of $P_{n0}$ (solid line)and $P_{nc}$(dashed line) with
$n$ for the {\bf quenched} limit ($r_0=4$)}
\end{figure}
\section{Conclusion.}
To conclude, in this paper we have constructed an effective
Hamiltonian  for the study of the dynamics of charged and neutral
scalar particles in an expanding background  FRW metric. Starting
from an O(4) sigma model through the inclusion of second order
quantum fluctuations we have shown that the Hamiltonian is quite
generic and in addition to applications to the formation of the
DCC \cite{qftdcc}, can also be used to study the enhancement of
low momentum charged and neutral particle modes in an FRW metric
suggesting an application to recent ideas of "preheating" in an
inflationary cosmological scenario. We have seen the appearance of
SU(1,1) symmetries leading to the presence of squeezed states in
their dynamics.  We find that in the quenched limit ( fast
expansion) the low momentum modes are enhanced significantly,
whereas in the adiabatic (slow expansion) limit, no such
enhancement occurs. This has been shown to be directly related to
the value of the squeezing parameter. The manifestation of this
difference shows up directly in the total neutral and charged
number distributions at zero momentum.

\section{Acknowledgements}
BB would like to thank the DST (India) and INSA (India) for
partial travel support for attending the International Congress of
Mathematical Physics (ICMP 2003) where this work was presented.
She would also like to thank Prof. Jean-Claude Zambrini of the
University of Lisbon for assistance and hospitality in lisbon. CM
would like to thank Prof.K.N. Pathak, Vice-Chancellor, Panjab
University for supporting his research. His research was actively
hindered by the University Grants Commission (India), under its
"Research Scientists" scheme. Hence this work was financed through
his salary as UGC Research Scientist "C" (Professor level!!)
graciously released by Panjab University.

\end{document}